\documentclass[letterpaper,twocolumn,prl,aps,superscriptaddress,amsmath,amssymb,floatfix]{revtex4-1}
\usepackage{mathptmx}
\DeclareMathAlphabet{\mathcal}{OMS}{cmsy}{m}{n}

\usepackage[latin9]{inputenc}
\usepackage{color}
\usepackage{verbatim}
\usepackage{float}
\usepackage{amsmath}
\usepackage{amssymb}
\usepackage{graphicx}
\usepackage{esint}
\usepackage[unicode=true,
 bookmarks=true,bookmarksnumbered=false,bookmarksopen=false,
 breaklinks=false,pdfborder={0 0 1},backref=false,colorlinks=true]
 {hyperref}
\hypersetup{
 linkcolor=magenta,urlcolor=blue,citecolor=blue,pdfstartview={FitH},hyperfootnotes=false}
\makeatletter

\usepackage{epsfig}\usepackage{amsfonts}

\usepackage{soul}

\makeatother

\begin{document}
\title{Integrated photonic nonreciprocal devices based on susceptibility-programmable medium}

\author{Yan-Lei Zhang}
\thanks{These two authors contributed equally to this work.}
\affiliation{CAS Key Laboratory of Quantum Information, University of Science
and Technology of China, Hefei, Anhui 230026, China}
\affiliation{CAS Center For Excellence in Quantum Information and Quantum Physics,
University of Science and Technology of China, Hefei, Anhui 230026, China}

\author{Ming Li}
\thanks{These two authors contributed equally to this work.}
\affiliation{CAS Key Laboratory of Quantum Information, University of Science
and Technology of China, Hefei, Anhui 230026, China}
\affiliation{CAS Center For Excellence in Quantum Information and Quantum Physics,
University of Science and Technology of China, Hefei, Anhui 230026, China}
\author{Xin-Biao Xu}
\affiliation{CAS Key Laboratory of Quantum Information, University of Science
and Technology of China, Hefei, Anhui 230026, China}
\affiliation{CAS Center For Excellence in Quantum Information and Quantum Physics,
University of Science and Technology of China, Hefei, Anhui 230026, China}
\author{Zhu-Bo Wang}
\affiliation{CAS Key Laboratory of Quantum Information, University of Science
and Technology of China, Hefei, Anhui 230026, China}
\affiliation{CAS Center For Excellence in Quantum Information and Quantum Physics,
University of Science and Technology of China, Hefei, Anhui 230026, China}
\author{Chun-Hua Dong}
\affiliation{CAS Key Laboratory of Quantum Information, University of Science
and Technology of China, Hefei, Anhui 230026, China}
\affiliation{CAS Center For Excellence in Quantum Information and Quantum Physics,
University of Science and Technology of China, Hefei, Anhui 230026, China}
\affiliation{Hefei National Laboratory, University of Science and Technology of China, Hefei 230088, China.}
\author{Guang-Can Guo}
\affiliation{CAS Key Laboratory of Quantum Information, University of Science
and Technology of China, Hefei, Anhui 230026, China}
\affiliation{CAS Center For Excellence in Quantum Information and Quantum Physics,
University of Science and Technology of China, Hefei, Anhui 230026, China}
\author{Chang-Ling Zou}
\email{clzou321@ustc.edu.cn}
\affiliation{CAS Key Laboratory of Quantum Information, University of Science
and Technology of China, Hefei, Anhui 230026, China}
\affiliation{CAS Center For Excellence in Quantum Information and Quantum Physics,
University of Science and Technology of China, Hefei, Anhui 230026, China}
\affiliation{Hefei National Laboratory, University of Science and Technology of China, Hefei 230088, China.}
\author{Xu-Bo Zou}
\email{xbz@ustc.edu.cn}
\affiliation{CAS Key Laboratory of Quantum Information, University of Science
and Technology of China, Hefei, Anhui 230026, China}
\affiliation{CAS Center For Excellence in Quantum Information and Quantum Physics,
University of Science and Technology of China, Hefei, Anhui 230026, China}
\affiliation{Hefei National Laboratory, University of Science and Technology of China, Hefei 230088, China.}
\date{\today}
\begin{abstract}
The switching and control of optical fields based on nonlinear optical effects are often limited to relatively weak nonlinear susceptibility and strong optical pump fields. Here, an optical medium with programmable susceptibility tensor based on polarizable atoms is proposed. Under a structured optical pump, the ground state population of atoms could be efficiently controlled by tuning the chirality and intensity of optical fields, and thus the optical response of the medium is programmable in both space and time. We demonstrate the potential of this approach by engineering the spatial distribution of the complex susceptibility tensor of the medium in photonic structures to realize nonreciprocal optical effects. Specifically, we investigate the advantages of chiral interaction between atoms and photons in an atom-cladded waveguide, theoretically showing that reconfigurable, strong, and fastly switchable isolation of optical signals in a selected optical mode is possible.
The susceptibility-programmable medium provides a promising way to efficiently control the optical field, opening up a wide range of applications for integrated photonic devices and structured optics.
\end{abstract}
\maketitle

\emph{Introduction.}- Precise control of optical medium has become increasingly
important for applications ranging from optical signal processing~\citep{almeida2004all,bi2011chip,kulheim2002rapid,Joannopoulos2011,Ruan2021}, imaging and microscopy~\cite{Gigan2022,Bertolotti2022}, biomedical tweezers~\cite{Moffitt2008}, and machine learning~\cite{Lin2018,Zhou2021}. The propagation and distribution of electromagnetic fields can be controlled by engineering the dielectric constants of materials~\cite{liberal2017near,shelby2001experimental,parazzoli2003experimental,smith2000negative,smith2004metamaterials,valentine2008three,ramakrishna2005physics,Yu2020,Forbes2021}, mainly by designing the geometric structure of dielectric materials to achieve a desired spatial distribution of refractive index. Previous micro- and nano-photonic techniques~\cite{Prasad2004,Elshaari2020} have enabled novel approaches for controlling optical fields, such as photonic crystal~\cite{istrate2006photonic,inoue2023self}, plasmonics~\cite{girard2008shaping}, metamaterials~\cite{engheta2007circuits}, and metasurfaces~\cite{li2022metasurface, liu2022programmable}.
The temporal modulation and dynamic reconfiguration of optical media is also of great importance~\cite{rotter2017light}. Rather than mechanically changing the structure geometry, nonlinear optical effects are typically used to modify the dielectric constant of materials through applied external fields~\cite{boyd2020nonlinear}. However, such approaches are usually limited to stringent phase matching conditions between the pump and signal fields and the weak nonlinear optical responses of materials, and these limitations lead to various challenges in practical optical devices, such as the magnetic-free nonreciprocal susceptibility of materials~\cite{Saleh1991,Yu2009,asadchy2020tutorial}. Therefore, optical media allowing for efficient programmable susceptibility in both spatial and temporal domains are highly demanded.

In the past decade, exotic phases of materials have been utilized for reconfigurable optical components. For example, spatial light modulators based on liquid crystals are widely applied~\cite{efron1994spatial}. Recently, phase-change materials are applied in reconfigurable nonvolatile metasurface~\cite{wu2021programmable,zhang2021electrically} and machine learning applications~\cite{feldmann2019all}. Alternatively, dilute atomic ensembles provide an excellent platform for realizing functional optical media~\cite{lin2019nonreciprocal, lu2021nonreciprocity, yang2019realization}. The couplings between light with transitions of atomic internal energy levels determine the susceptibility~\citep{scully1991enhancement,bons2016quantum}, with its real and imaginary parts corresponding to dispersion and absorption, respectively. Therefore, by effectively dressing the internal energy levels of atoms with a near-resonant control optical field, the susceptibility of atomic media could be modulated. In particular, nonreciprocal signal propagation has been realized based on the spatiotemporal modulation of the control field, i.e., the \textquotedblleft moving\textquotedblright{}
Bragg mirror~\citep{horsley2013optical}. Even the quantum property of transmitted and reflected signals have been proposed by quantum metasurfaces made of atom arrays~\cite{bekenstein2020quantum}.

\begin{figure}
\includegraphics[width=0.5\textwidth]{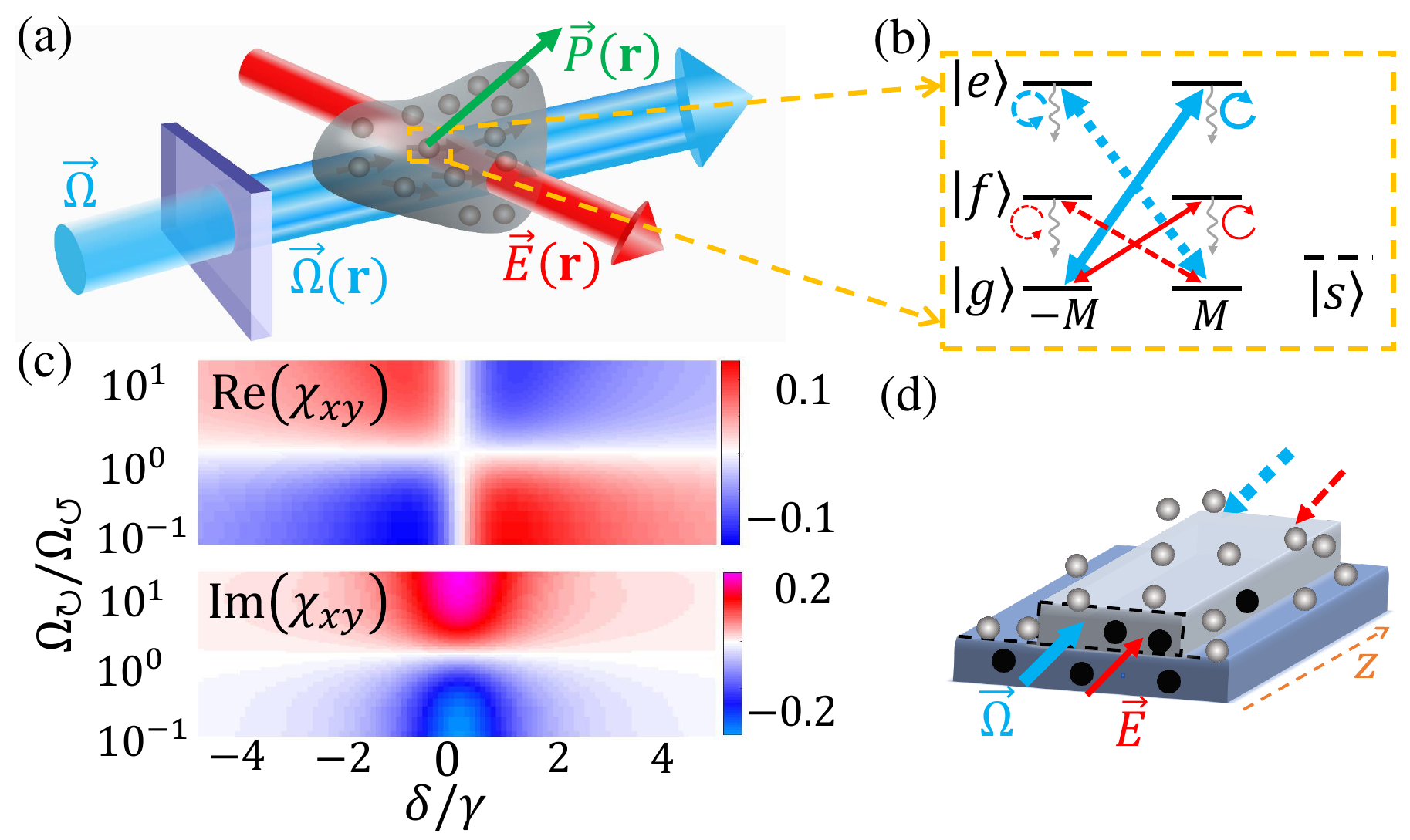}
\caption{(a) Atomic polarization $\protect\overrightarrow{P}$ for the signal field $\protect\overrightarrow{E}$ under the control field $\protect\overrightarrow{\Omega}$ modulated by the spatial light modulator. (b) Energy levels diagram. The control field (blue arrow) stimulates the transition $\left|e\right\rangle \leftrightarrow\left|g\right\rangle$, the signal field (red arrow) stimulates the transition $\left|f\right\rangle \leftrightarrow\left|g\right\rangle $, and $\left|s\right\rangle$ is an auxiliary level to turn off the interaction between light and the atom. (c) The signal susceptibility as a function of the ratio $\Omega_{\circlearrowright}/\Omega_{\circlearrowleft}$ and the detuning $\delta$: the dispersion $\mathrm{Re\left(\chi_{xy}\right)}$ (top) and the absorption $\mathrm{Im\left(\chi_{xy}\right)}$ (bottom). (d) The general schematic of the atom-cladded waveguide (white atoms) and the solid-state emitter-doped waveguide (black atoms).}
\label{Fig1}
\end{figure}

In this Letter, we propose susceptibility-programmable media (SPMs) for manipulating optical field and realizing novel integrated photonic devices. By stimulating atoms into different long-lived ground energy levels, the optical responses of atoms can be tailored, and the susceptibility of atomic media can be programmed. The scheme can be easily adapted in hybrid photonic systems, such as atom-cladded waveguide~\cite{corzo2019waveguide,zektzer2021nanoscale} and nanofiber~\cite{ma2022composite} or solid state emitter-doped photonic structures~\cite{Gritsch2022,WangSihao2022} without requiring either complex external controls or fabricating complex structures, SPMs have the potential for classical and quantum information processing applications, such as mode conversion, tunable optical interference, chirality detection.

\emph{Susceptibility-programmable medium.}- Figure~\ref{Fig1}(a) illustrates a general setup for modulating the propagation of optical signals by a control laser field $\overrightarrow{\Omega}$. At a location $\bold{r}$, the polarization of the medium due to the signal field $\overrightarrow{E}(\bold{r})$ can be written as $\overrightarrow{P}(\bold{r})=\epsilon_{0}\overleftrightarrow{\chi}(\bold{r},\overrightarrow{\Omega}(\bold{r}))\cdot\overrightarrow{E}(\bold{r})$~\citep{momeni2019generalized,kauranen1997quantitative,boyd2020nonlinear}, where $\epsilon_{0}$
is the vacuum permittivity, and $\overleftrightarrow{\chi}(\bold{r},\overrightarrow{\Omega}(\bold{r}))$ is
the local susceptibility tensor of the medium under a control field $\overrightarrow{\Omega}(\bold{r})$ modulated by a spatial light modulator. $\overleftrightarrow{\chi}$ is a second-rank tensor of an anisotropic material, which can be realized with an atomic medium, whose populations on ground state Zeeman energy levels could be polarized by the control field. Figure~\ref{Fig1}(b) shows a simplified energy level structure of an atom, where $\left|e\right\rangle$, $\left|f\right\rangle$, and $\left|g\right\rangle$ represent a three-level structure, and $-M,...,m,m+1,...,M$ are Zeeman levels. The signal and control lights are near-resonant with $g-f$ and $g-e$ transitions, respectively. An idle energy level $\left|s\right\rangle$ is also introduced so that the atom at this state is transparent for the control and signal fields to turn-off the interactions.
Without loss of generality, the Zeeman levels are simplified to two fine levels $m=\pm M$~\cite{wang2022self,Yang2023}, and the input light traveling along the atom quantization axis ($z$ direction) can be decomposed into right-polarized  $\circlearrowright=\overrightarrow{e_{x}}+i\overrightarrow{e_{y}}$
and left-polarized $\circlearrowleft=\overrightarrow{e_{x}}-i\overrightarrow{e_{y}}$ components, that is, the signal and control fields can be decomposed as $\overrightarrow{E}\rightarrow E_{\circlearrowright}, E_{\circlearrowleft}$ and $\overrightarrow{\Omega}\rightarrow \Omega_{\circlearrowright}, \Omega_{\circlearrowleft}$ (for a more general description of the SPM, see~\cite{sm}).
Therefore, the corresponding Hamiltonian ($\hbar=1$) of the system can be described as:
\begin{eqnarray}
H & = & \sum_{k=1}^{N}\sum_{j=g,f,e}\omega_{j,k}\left(\sigma_{jj,k}^{-M,-M}+\sigma_{jj,k}^{M,M}\right)\nonumber \\
 &  & +\left(\Omega_{\circlearrowright,k}\sigma_{ge,k}^{-M,M}+\Omega_{\circlearrowleft,k}\sigma_{ge,k}^{M,-M}\right)e^{i\omega_{\Omega}t}+\mathrm{h.c.}\nonumber \\
 &  & +\left(E_{\circlearrowright,k}\sigma_{gf,k}^{-M,M}+E_{\circlearrowleft,k}\sigma_{gf,k}^{M,-M}\right)e^{i\omega_{s}t}+\mathrm{h.c.}\,
 \label{Eq2}
\end{eqnarray}
where $N$ is the atomic number, $\sigma_{jj',k}^{\pm M,\mp M}=\left|j,\pm M\right\rangle_{k} \left\langle j',\mp M\right|$ with $j,j'\in \left \{g,f,e\right \}$, and $\omega_{\Omega}$ and $\omega_{s}$ are the frequencies of the control and signal fields, respectively. Here, coherent atom-atom interaction is neglected due to low atomic density, and the broadening of atomic transitions due to the atomic collision is considered. When the signal field is very weak compared to the control field, i.e., $|\overrightarrow{E}|\ll|\overrightarrow{\Omega}|$, the control light can polarize the atom ground states toward level $M$ or $-M$, and the atomic ground state population is determined by the control field.

Without loss of generality, we assume that the atoms are identical for convenience when discussing the susceptibility. For the input signal traveling along the atom quantization axis ($z$ direction), the second-rank susceptibility tensor~\cite{sm} reads
\begin{equation}
\overleftrightarrow{\chi}=\frac{1}{2}\left[\begin{array}{cc}
\chi_{xx} & -i\chi_{xy}\\
i\chi_{xy} & \chi_{yy}
\end{array}\right],
\end{equation}
where $\chi_{xx}=\chi_{yy}=\chi_{\circlearrowright}+\chi_{\circlearrowleft}$, $\chi_{xy}=\chi_{\circlearrowleft}-\chi_{\circlearrowright}$, with circular susceptibilities~\cite{sm}:
\begin{align}
\chi_{\circlearrowright}
& \approx  \frac{i\rho_a\left|\mu_{fg}\right|^{2}}{\left(i\delta+\gamma\right)\varepsilon_{0}\hbar} \frac{\left(\gamma^2+\Delta^2+\Omega_{\circlearrowright}^{2}\right)\Omega_{\circlearrowleft}^{2}}{\left(\gamma^2+\Delta^2\right)\left(\Omega_{\circlearrowright}^{2}+\Omega_{\circlearrowleft}^{2}\right)+4\Omega_{\circlearrowright}^{2}\Omega_{\circlearrowleft}^{2}},\label{Eq3}\\
\chi_{\circlearrowleft}
& \approx  \frac{i\rho_a\left|\mu_{fg}\right|^{2}}{\left(i\delta+\gamma\right)\varepsilon_{0}\hbar}  \frac{\left(\gamma^2+\Delta^2+\Omega_{\circlearrowleft}^{2}\right)\Omega_{\circlearrowright}^{2}}{\left(\gamma^2+\Delta^2\right)\left(\Omega_{\circlearrowright}^{2}+\Omega_{\circlearrowleft}^{2}\right)+4\Omega_{\circlearrowright}^{2}\Omega_{\circlearrowleft}^{2}}.\label{Eq4}
\end{align}
Here, $\delta=\omega_{g}-\omega_{f}+\omega_{s}$ and $\Delta=\omega_{g}-\omega_{e}+\omega_{\Omega}$ are detunings with respect to the transitions, $\gamma$ is the decay rate of $\left|e\right\rangle$ and $\left|f\right\rangle$, $\rho_a$ is the atomic density, $\mu_{fg}$ the electric dipole moment, and $\varepsilon_{0}$ is the vacuum permittivity. The susceptibility of the signal is determined by the control field $\Omega_{\circlearrowright\left(\circlearrowleft\right)}$.
In particular, when $\Omega_{\circlearrowright}\neq\Omega_{\circlearrowleft}$, we have the chiral susceptibility $\chi_{xy}\neq0$ and the medium exhibiting nonreciprocity because circular dichroism or birefringence for $\chi_{\circlearrowright}\neq\chi_{\circlearrowleft}$~\cite{asadchy2020tutorial}.

\begin{figure*}
\includegraphics[width=1\textwidth]{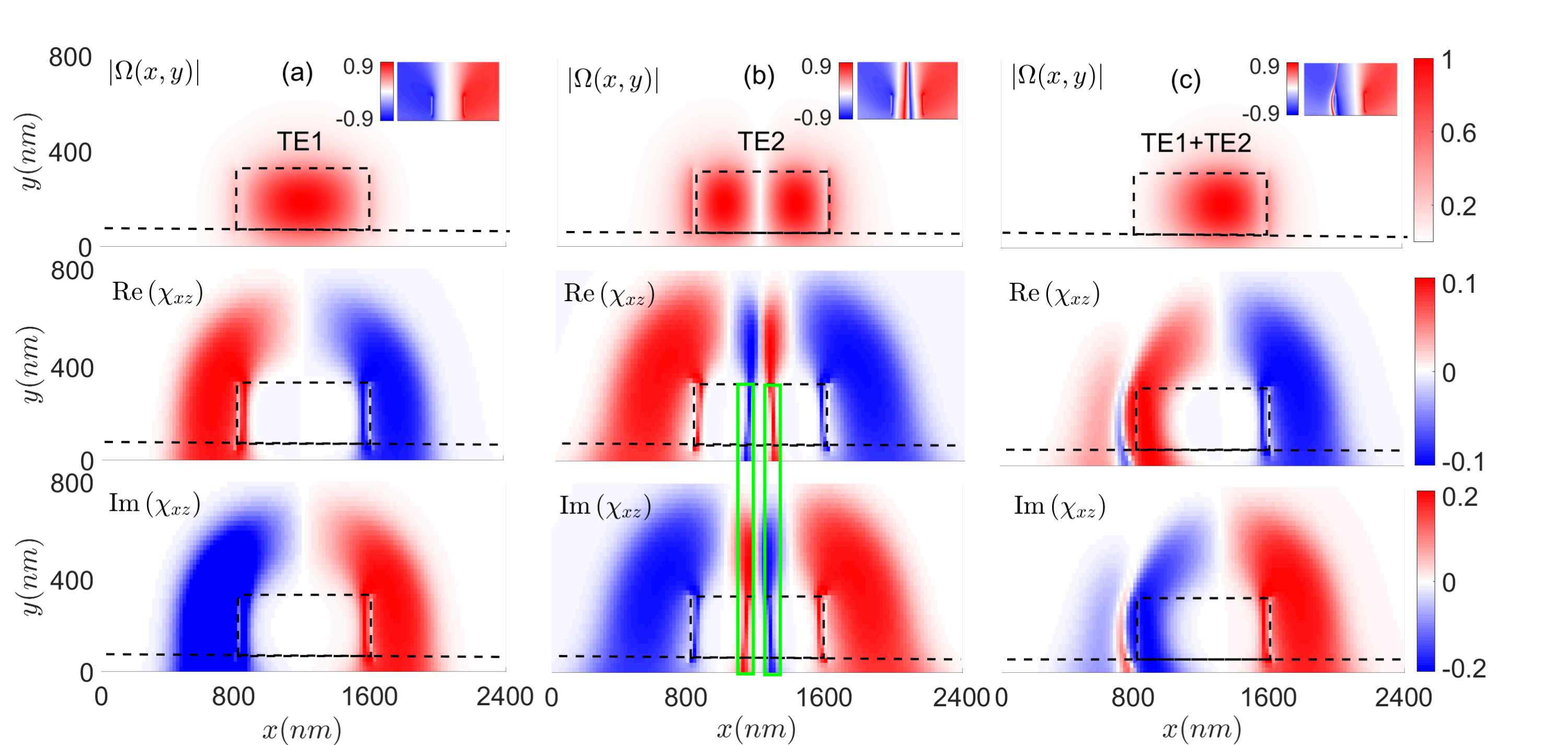}\caption{Spatial distributions of optical field $|\Omega (x,y)|$ (top), the induced dispersion $\mathrm{Re}\left(\chi\right)$ (middle) and absorption $\mathrm{Im}\left(\chi\right)$ (bottom) for the signal field at the cross-section of the waveguide, with the detuning $\delta=0.5\gamma$. (a)-(c) are the results for the fundamental mode (TE1), the second-order mode (TE2), and the superposition of TE1 and TE2, respectively. The black dotted line is the boundary of waveguide, and the inset at the upper right corner represents the chirality $(\Omega_{\circlearrowright}-\Omega_{\circlearrowleft})/(\Omega_{\circlearrowright}+\Omega_{\circlearrowleft})$. The green boxes in (b) show that the susceptibility has a sharp change in space, which is about 10\,nm. The parameters are $\gamma=10$ MHz, $\rho_{a}=4\times10^{18}/\mathrm{m}^{3}$, $\mu_{fg}=1.4\times10^{-29}\mathrm{C.m}$, $\hbar=1.055\times10^{-34}\mathrm{J.s}$, $\varepsilon_{0}=8.85\times10^{-12}\mathrm{F/m}$, $J=10^{-4}\gamma$ and $\Omega=10^{2}E\simeq\gamma$. }
\label{Fig2}
\end{figure*}

Figure~\ref{Fig1}(c) shows the dependence of $\chi_{xy}$ on the control field ratio $\Omega_{\circlearrowright}/\Omega_{\circlearrowleft}$ and detuning are numerically studied, with $\gamma=10$ MHz, $\rho_{a}=1.26\times10^{19}/\mathrm{m}^{3}$~\cite{ketterle1993high,sebby2005high}, $\Omega_{\circlearrowleft}=0.3 \gamma$ and $|\overrightarrow{E}|\ll\gamma, |\overrightarrow{\Omega}|$. The results implies a significant change of the effective refractive index $n_{\mathrm{eff}}=\sqrt{1+\chi}$~\cite{boyd2020nonlinear, scully1999quantum} of the medium by $10\%$ through control fields. Therefore, the susceptibility could be efficiently controlled by only engineering the control field intensity distribution without requiring the phase-matching condition between the control and signal. Additionally, unlike the optical dressing approaches~\cite{horsley2013optical,zhang2018thermal}, the phase coherence between the two fields is not required. Therefore, the SPM holds advantages including the high spatial and temporal resolution in dynamical reconfiguration of the material, the robustness against detuning and phase incoherence of the control field~\cite{Hu2021}, and the inhomogeneous broadening of atoms, as demonstrated later.

\emph{Programmable nonreciprocal device.}- Figure~\ref{Fig1}(d) shows an example of realizing magnetic-free nonreciprocal optical devices by employing the SPM. Atom-like emitters, such as hot atom ensembles~\cite{wu2010slow}, trapped cold atoms~\cite{ohtsu1999near,ma2022composite}, deposited layers of solid state single emitters~\cite{wu2021two}, ion-doped dielectrics~\cite{lodahl2015interfacing}, and solid state emitter-doped photonic structures~\cite{Gritsch2022,WangSihao2022}, could be introduced to either on the top of waveguide or doped into the waveguide. We note that the atoms can be randomly distributed around or inside the waveguide without special ordering. The combination of SPM and tightly confined optical modes in photonic structures has the advantages of: (1) laterally confined optical fields posses local chirality~\cite{zektzer2019chiral}, which is essential for realizing chiral susceptibility. (2) atom-photon interaction is greatly enhanced, thus allowing the realization of high-performance devices with a small footprint and relatively low atom density.

In contrast to the optical polarization in free space, the input $\overrightarrow{\Omega}\left(x,y\right)$ could polarize atoms according to the chirality of the evanescent field~\cite{zektzer2019chiral} of a SiN waveguide (width $w=800\,$nm and thickness $h=360\,$nm). Through numerical simulation, we obtained the electric field distributions of the waveguide: including $\Omega_{\circlearrowright}\left(x,y\right)$, $\Omega_{\circlearrowleft}\left(x,y\right)$, $E_{\circlearrowright}\left(x,y\right)$, and $E_{\circlearrowleft}\left(x,y\right)$, where we redefine $\circlearrowright=\overrightarrow{e_{x}}+i\overrightarrow{e_{z}}$ and $\circlearrowleft=\overrightarrow{e_{x}}-i\overrightarrow{e_{z}}$. Figure~\ref{Fig2}(a) displays the distributions of the fundamental transverse electric (TE1) mode at the cross section, with the top panel and its inset showing the normalized field amplitude $|\Omega_{x,y}|$ and the field chirality $(\Omega_{\circlearrowright}-\Omega_{\circlearrowleft})/(\Omega_{\circlearrowright}+\Omega_{\circlearrowleft})$, respectively. As expected, the mode's evanescent field on both sides of the waveguide exhibits chirality. The corresponding response of signal field is determined by the chiral susceptibility $\chi_{xz}$, which can be obtained by numerically solving the master equation, under the assumption of uniformly distributed atoms and considering the decay of excited states and the relaxation of ground states. The chiral susceptibilities are summarized by the middle and bottom panels in Fig.~\ref{Fig2}(a), showing drastically varying of $\chi_{xz}$ in spatial due to the tightly confined control field and its chirality [inset of Fig.~\ref{Fig1}(a)].

Compared with Fig. \ref{Fig2}(a), the distribution of the chirality is distinct for second-order mode (TE2) in Fig.~\ref{Fig2}(b), as is the susceptibility. In particular, there are subwavelength features labelled by the green boxes, whose mechanism is similar to that of stimulated emission depletion (STED) microscopy~\cite{Hell1994}. The $\chi$ of the atoms is a nonlinear function against $\Omega$ [Eqs.~(\ref{Eq3}) and (\ref{Eq4})]. Employing the superposition of different modes, the spatial distribution of the susceptibility could be further engineered. As an example, the distribution of the control field as a superposition of TE1 and TE2 $\Omega_{\circlearrowright\left(\circlearrowleft\right)}\left(x,y\right)=\xi\Omega_{\circlearrowright\left(\circlearrowleft\right)}^{1}\left(x,y\right)+\sqrt{1-\xi^{2}}\Omega_{\circlearrowright\left(\circlearrowleft\right)}^{2}\left(x,y\right)$ could be precisely adjusted by controlling $\xi$ with $0\leq\xi\leq1$. In Fig.~\ref{Fig2}(c), the results with the parameter $\xi=1/\sqrt{2}$ are presented, showing significantly asymmetric distributions of $\chi_{xz}$.

In addition, as we can infer from Fig.~\ref{Fig1}(c), the spatial distribution of the susceptibility can also be tuned by adjusting the control laser detuning $\Delta$, the signal detuning $\delta$, and the intensities of the control $\Omega$ and the signal $E$. If the assisted level $\left|s\right\rangle$ is also employed, a more complicated susceptibility distribution can be designed as needed. If the atoms are prepared to the state $\left|s\right\rangle$, the atoms are transparent to the signal which is far off-resonance with the transitions between $\left|s\right\rangle$ and other energy levels. Therefore, a tunable scatter in a microcavity can be realized by selectively pumping atoms out from $\left|s\right\rangle$ at certain positions. Through controlling the distribution of the localized scattering points, mode conversion and atomic grating can be realized~\cite{sm}. We note that the waveguide also supports the TM (transverse magnetic) modes, and similar susceptibility engineering can be realized while the spatial distribution shows different features~\cite{sm}.  Based on the results above, we conclude that we can realize SPMs by employing the high-order modes of the waveguide and other adjustable parameters.

\begin{figure}
\includegraphics[width=0.5\textwidth]{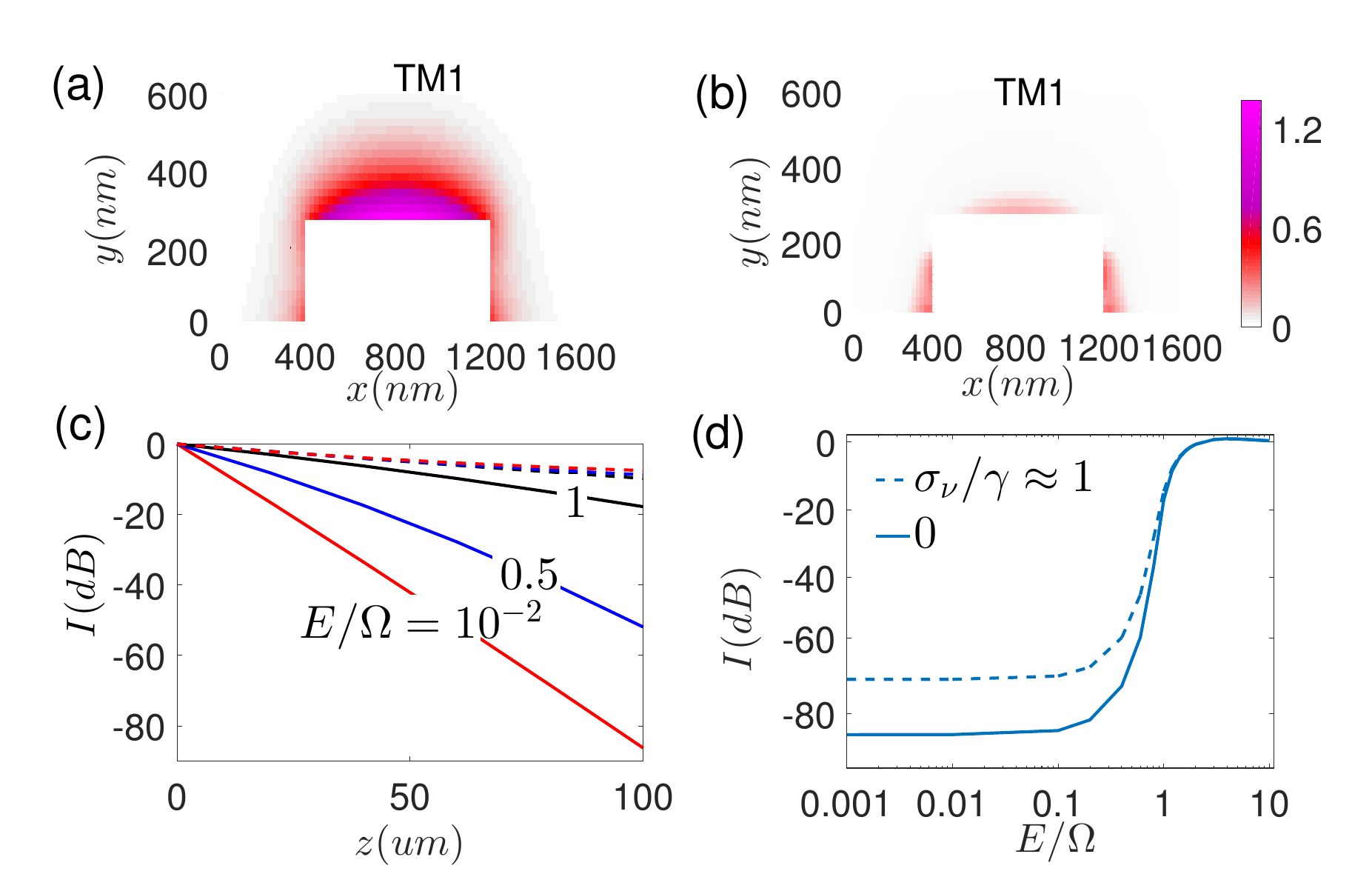}
\caption{(a) and (b) The spatial distribution of the effective absorption of the signal field per unit area for the control light (TM1 mode) propagating in the opposite (a) and same (b) directions, respectively. (c) Optical isolation $I$ as a function of waveguide length $z$ for different control field strengths: $E=\Omega$ (black line), $\Omega/2$ (blue line), and $10^{-2}\Omega$ (red line). The dashed lines correspond to the insertion loss. (d) Nonreciprocity $I$ as a function of the ratio $E/\Omega$, comparing the cases with (dashed line, $\sigma_{\nu}/\gamma\simeq1$) and without (solid line) inhomogeneous broadening. Other parameters are the same as in Fig.~\ref{Fig2}, and the detuning of signal is zero.
}
\label{Fig4}
\end{figure}

\emph{Optical isolation.}- Nonreciprocal or unidirectional devices are indispensable in anti-noise sensing and back-action-immune communications or information processing. Based on the above analysis, the field chirality of the control field induces chiral susceptibility  for realizing the nonreciprocal propagation of the signal, and the nonreciprocal susceptibility for unidirectional transmission is optimal when the atoms are around the waveguide, employing the TM1 mode of the waveguide. Here we want to emphasize that the nonreciprocity of signal transmission from one port to another port at the same TM1 mode, and the Lorentz reciprocity is broken due to the asymmetric susceptibility induced by the unidirectional control light~\cite{jalas2013and}. The propagation of the signal field amplitude ($E\left(z\right)$) along the waveguide follows the transmission equation~\cite{scully1999quantum}
\begin{equation}
\frac{dE\left(z\right)}{dz}=-k_{s}E(z)\iint \rho(x,y)\overrightarrow{u_s}\left(x,y\right)\overleftrightarrow{\chi}(x,y,z)\overrightarrow{u_s}^{\dagger}\left(x,y\right)dxdy,\nonumber
\end{equation}
where $k_{s}=2\pi/\lambda$ is the wave vector of the signal light, $\overrightarrow{u_s}\left(x,y\right)$ is the normalized field distribution at the cross section, $\rho(x,y)$ is the atomic density, and $\overleftrightarrow{\chi}(x,y,z)$ is the susceptibility.
When considering the potential isolation of the signal, i.e., the difference in signal attenuation for the forward and backward directions, the imaginary part of the integral matters. Assuming atoms are uniformly distributed around the waveguide, the spatial distribution of the effective absorption of the signal field per unit area is numerically solved, and the results for counter-propagating and co-propagating signal and control lights are plotted in Figs.~\ref{Fig4}(a) and (b), respectively. When the signal and the control light are in the opposite direction, the effective absorption is mainly distributed on the upper surface of the waveguide, where the field chirality is significant and meanwhile the evanescent field is strongest. It is estimated that the effective refractive index $n_{\mathrm{eff}}=\sqrt{1+\chi}<0.03$ for low atomic density $\rho_{a}=4\times10^{18}/\mathrm{m}^{3}$, which means that the mode largely remains unchanged. In contrast, when the signal light transmits along the same direction as the control light, the absorption is suppressed by around one order of magnitude. By also considering the varying control field strength along the waveguide due to the absorption of atoms, the transmitted signal at different propagation directions for a given waveguide length $z$ is numerically solved, and the corresponding isolation ratio is obtained. The huge difference in the effective absorption leads to the nonreciprocal transmission of the signal light, as shown in Fig.~\ref{Fig4}(c). Here, isolation is defined as $I=10\log \left|E_{o}\left(z\right)/E\left(0\right)\right|^2-10\log \left|E_{s}\left(z\right)/E\left(0\right)\right|^2$, and the subscripts $o$ and $s$ indicate whether the signal light and the control light are transmitted in the opposite direction or in the same direction. Comparing different relative ratios of the input signal and control field strength, excellent isolation of $20~\mathrm{dB}$ can be achieved with a waveguide length less than $50\,\mathrm{\mu m}$ even when the signal is comparable to the control ($E=\Omega/2$), and the corresponding insertion loss is only about $4~\mathrm{dB}$. The isolation as a function of $E/\Omega$ is shown by the solid line in Fig.~\ref{Fig4}(d), and we conclude that the performance of nonreciprocity is still maintained with a wide range of light intensities.

For practical experiments, the transition frequencies of atoms in an ensemble might be broadened due to the Doppler effect, local stray fields, or mechanical strains in solids. Such inhomogeneous broadening of atoms can be described by a Gaussian distribution $D\left(\nu\right)=e^{-\nu^{2}/\sigma_{\nu}^{2}}/\left(\sigma_{\nu}\sqrt{\pi}\right)$, with $\nu$ is the extra frequency shift of transitions and $\sigma_{\nu}$ is the standard deviation of the shifts. Therefore, the overall susceptibility can be rewritten as $\overleftrightarrow{\chi}(x,y,z)=\int_{-\infty}^{\infty}\overleftrightarrow{\chi}(x,y,z,\nu)D\left(\nu\right)d\nu$.Then, the effect of inhomogeneous broadening with $\sigma_{\nu}/\gamma\simeq1$ is numerically investigated, as shown by the dotted line in Fig.~\ref{Fig4}(d). Comparing the two curves, the performance of the isolation is weakened but still allows an ultra-high isolation when $E/\Omega\ll1$, which clearly displays the robustness against the inhomogeneous broadening of atoms.

\begin{figure}
\includegraphics[width=0.5\textwidth]{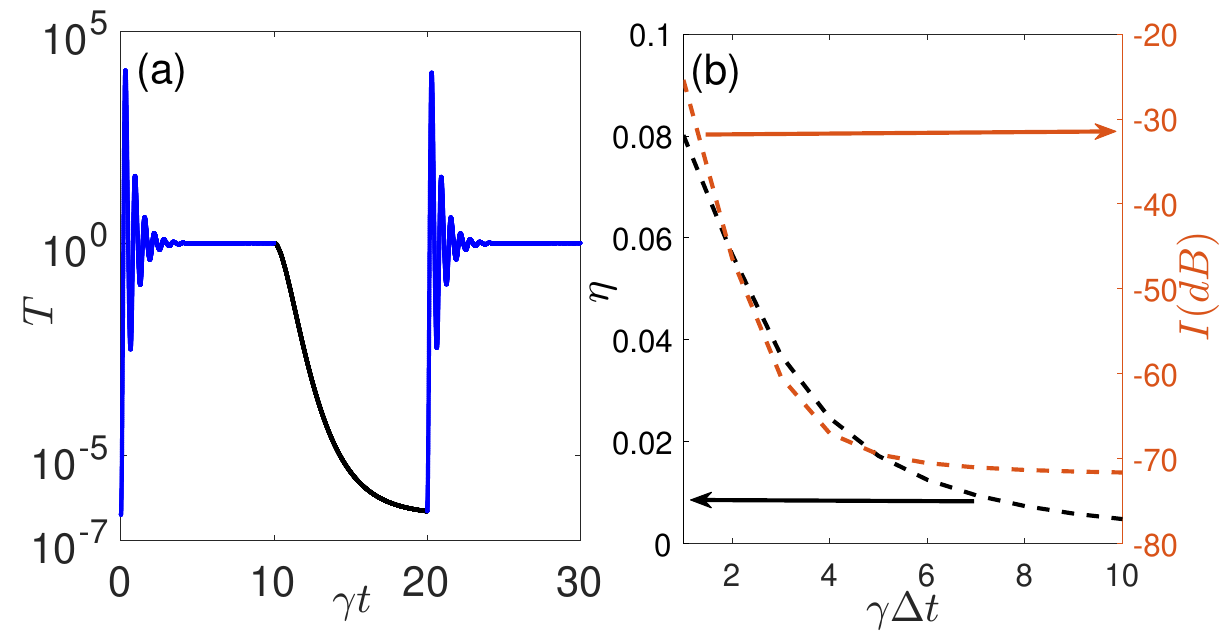}
\caption{(a) The dynamical process for the optical isolation, which shows the high speed switching that is proportional to atomic decay rate $\gamma$. The transmission is defined as $T_{o,s}=\left|E_{o,s}\left(z\right)/E\left(0\right)\right|^2$. The blue line is $T_{s}$, where  the signal light
and the control light are transmitted in the same direction and the black line means the opposite transmission $T_{o}$.
(b) The signal distortion for a pulse signal $E\left(t\right)\propto\exp\left(-t^{2}/\left|\Delta t\right|^{2}\right)$, where the distortion degree $\eta$ (the black line) is inversely proportional to the width of the pulsed signal and while the isolation $I$ is increased with the increasing of $\Delta t$.
}
\label{Fig5}
\end{figure}

In Fig.~\ref{Fig5}, we investigate the dynamical process and the distortion of the signal. The numerical results show that the switching speed is proportional to the atomic decay rate $\gamma$, indicating that the system requires a duration of about $1/\gamma$ to reach stability, as depicted in Fig.~\ref{Fig5} (a). Furthermore, we observe an interesting temporal amplification effect of the signal, which is attributed to population inversion due to the impulse response of SPM, which appeals further exploration of the programmable function. We have performed numerical simulations to investigate the extent of signal distortion and its dependence on the pulse width and the properties of the atomic medium in Fig.~\ref{Fig5}(b). Our numerical results demonstrate that the signal light experiences a very small distortion less than $1\%$ when passing through the waveguide with high isolation $I<-70$ dB, where $\gamma\Delta t>7$. For a more detailed discussion about the switching speed and the distortion, see~\cite{sm}.

\emph{Conclusion.}- Based on the mechanism by which the states of atom or atom-like structures can be reconfigured by input optical fields, a susceptibility programmable medium (SPM) is proposed and theoretically investigated. The spatial distribution and elements of the susceptibility tensor of the medium are highly controllable by certain optical control fields. For example, by employing the inherent field chiral property of confined optical modes in photonic waveguides, spatial-dependent chiral susceptibility is achieved with SPM, and nonreciprocal propagation of the signal field in the same waveguide is allowed, and we can further expand more applications, including the mode conversion, the tunable optical interference, and the chirality detection (see ~\cite{sm}). The SPM can also be generalized to other novel functional photonic devices, including the nonreciprocal optical components, optical switches, and reconfigurable metasurfaces.

\begin{acknowledgments}
This work was funded by the National Key R\&D Program (Grant No. 2021YFA1402004), the National Natural Science Foundation of China (Grants 92265108, U21A20433, U21A6006, 12104441, 12061131011, and 92265210), and the Natural Science Foundation of Anhui Province (Grant Nos. 2108085MA17 and 2108085MA22). This work was also supported by the Fundamental Research Funds for the Central Universities and USTC Research Funds of the Double First-Class Initiative. The numerical calculations in this paper have been done on the supercomputing system in the Supercomputing Center of the University of Science and Technology of China. This work was partially carried out at the USTC Center for Micro and Nanoscale Research and Fabrication.
\end{acknowledgments}


%

\end{document}